\documentclass[10pt]{article} 
\usepackage{caption}

\usepackage{subcaption}
\usepackage[utf8]{inputenc}

\usepackage{amssymb} 
\usepackage{amsfonts}
\usepackage{amsmath}
\usepackage{graphicx}
\usepackage{indentfirst}
\usepackage{cite}

\usepackage{amsthm}
\theoremstyle{remark}
\newtheorem{theorem}{Theorem}
\newtheorem{corollary}{Corollary}

\usepackage{mathtools}

\usepackage{enumerate}
\usepackage{url}

\usepackage{lmodern}
\usepackage{newpxtext}

\usepackage[symbol]{footmisc}
\renewcommand{\thefootnote}{\fnsymbol{footnote}}

\usepackage{tikz}
\definecolor{darkblue}{rgb}{0.15,0.35,0.55}
\definecolor{reddish}{rgb}{.8, 0.2, 0.2}
\usepackage[pdfpagelabels, linktocpage=true, hypertexnames=false]{hyperref}

\hypersetup{
colorlinks=true,
citecolor=darkblue,
linkcolor=reddish,
urlcolor=darkblue,
pdfauthor={},
pdftitle={},
pdfsubject={}
}

\long\def\ca#1\cb{} 

\newcommand{\becs}{\begin{cases}}
\newcommand{\bem}{\begin{matrix}}

\newcommand{\bra}[1]{\langle#1|}

\newcommand{\bsk}{\bigskip }

\newcommand{\dya}[1]{|#1\rangle\langle#1|}

\newcommand{\encs}{\end{cases}}
\newcommand{\enm}{\end{matrix}}

\newcommand{\ket}[1]{|#1\rangle }

\newcommand{\ot}{\otimes }

\newcommand{\Tr}{{\rm Tr}}

\newcommand{\BC}{{\mathcal B}}
\newcommand{\CC}{{\mathcal C}}
\newcommand{\DC}{{\mathcal D}}

\newcommand{\HC}{{\mathcal H}}
\newcommand{\IC}{{\mathcal I}}

\newcommand{\QC}{{\mathcal Q}}

\newcommand{\TC}{{\mathcal T}}

\newcommand{\al}{\alpha }
\newcommand{\bt}{\beta }
\newcommand{\gm}{\gamma }

\newcommand{\Dl}{\Delta }
\newcommand{\ep}{\epsilon}

\newcommand{\lm}{\lambda }

\newcommand{\sg}{\sigma }

\newcommand{\om}{\omega }

\newcommand\blfootnote[1]{%
  \begingroup
  \renewcommand\thefootnote{}\footnote{#1}%
  \addtocounter{footnote}{-1}%
  \endgroup
}

\begin{document}

\begin{center}
{\bf \large Entropic singularities give rise to quantum transmission}\bsk

Vikesh Siddhu$^{1,2*}$~\blfootnote{$^1$~Department of Physics, Carnegie
Mellon University, Pittsburgh, Pennsylvania 15213, U.S.A, 
$^2$~JILA, University of Colorado/NIST, Boulder, CO 80309, U.S.A~(current address); $^*$email:
\href{mailto:vsiddhu@protonmail.com}{\texttt{vsiddhu@protonmail.com}}} \bsk

Date: October 1, 2021
\end{center}

\begin{abstract}
When can noiseless quantum information be sent across noisy quantum devices?
And at what maximum rate? These questions lie at the heart of quantum
technology, but remain unanswered because of non-additivity--- a fundamental
synergy which allows quantum devices (aka quantum channels) to send more
information than expected.  Previously, non-additivity was known to occur in
very noisy channels with coherent information much smaller than that of a
perfect channel; but, our work shows non-additivity in a simple low-noise
channel. Our results extend even further.  We prove a general theorem
concerning positivity of a channel's coherent information. A corollary of this
theorem gives a simple dimensional test for a channel's capacity.  Applying
this corollary solves an open problem by characterizing all qubit channels
whose complement has non-zero capacity. Another application shows a wide class
of zero quantum capacity qubit channels can assist an incomplete erasure
channel in sending quantum information. These results arise from introducing
and linking logarithmic singularities in the von-Neumann entropy with quantum
transmission: changes in entropy caused by this singularity are a mechanism
responsible for both positivity and non-additivity of the coherent information.
Analysis of such singularities may be useful in other physics problems.
\end{abstract}

\section{Introduction}

Entropy is fundamental. As a measure of complexity in a statistical
distribution, entropy is widely used in learning theory~\cite{SeifHafeziEA21,
PaulusGeyerEA90}, economics~\cite{GeorgescuRoegen71, Kummel11}, and
cryptography~\cite{Cachin97}. In physics, entropy usually quantifies disorder.
It is used to express laws of thermodynamics~\cite{Clausius51, Kardar07,
Jarzynski11}, explore the nature of black holes~\cite{Bekenstein73, Hawking76,
Wald01}, and study a variety of other physical phenomenon~\cite{Frenkel99,
CrossHohenberg93, AsorBennunShaulEA17, BaxaHaddadianEA14, ColesBertaEA17,
EisertCramerEA10, PresseGhoshEA13}.  Advances in understanding mathematical and
computational properties of entropy~\cite{Wehrl78, LiebRuskai73,
AvineryKornreichEA19, LezonBanavarEA06} have opened the doors for deeper
insights in physics and many other areas of study.

One area where entropy provides key insights is information
science~\cite{Shannon48}. The Shannon entropy not only quantifies
the amount of classical information in a source, it also plays a fundamental
role in answering a key practical question: when, and at what maximum rate can
classical information be sent across noisy communication channels? The maximum
Shannon entropy common between a channel's input and output, called the channel
mutual information $C^{(1)}$, gives an achievable rate at which error
correcting codes can recover noiseless information sent across many uses of a
noisy channel.  

The channel mutual information satisfies a crucial property, {\em additivity}:
the channel mutual information for two channels used together is the sum of
each. This additivity ensures that the channel capacity $C$, defined as the
best possible achievable rate, simply equals the channel mutual information
$C^{(1)}$. More remarkably, additivity implies that the channel capacity
completely specifies a classical channel's ultimate ability to send
information.  These implications are not only fundamental to our understanding
of noisy classical information but also critical to the use of channel capacity
as a benchmark for error correcting codes. These codes are essential for
storing and sending noiseless classical information across noisy channels
~\cite{CoverThomas01, MacKay02}.

The physical world is not classical but quantum mechanical. It contains quantum
information which is strikingly different from its classical
counterpart~\cite{WoottersZurek82,BennettBrassard14, BennettBrassardEA93,
KumarPatiBraunstein00}. 
In practice, noisy quantum devices carry quantum information. These
devices, which may send, store, or process information, are modelled
mathematically by completely positive trace preserving maps, also called
(noisy)~quantum channels.
While quantum information can be extremely useful for computing and
communication, it is notoriously error prone. 
Consequently, there are both fundamental and practical reasons to understand
when and at what maximum rate can noiseless quantum information be stored,
processed, or sent across noisy quantum channels~\cite{BennettDiVincenzoEA96,
BennettShor98,Lloyd97, BarnumKnillEA00, Shor02a, CaiWinterEA04, Devetak05}.
Despite dedicated efforts, there is no satisfactory answer to this basic
question.  The key reason behind this unsatisfactory state of affairs is
non-additivity in the quantum analog~\cite{SchumacherNielsen96,
BarnumNielsenEA98}, $\QC^{(1)}$, of the channel mutual information $C^{(1)}$:
for two noisy quantum channels $\BC_1$ and $\BC_2$ used in parallel, the
channel coherent information $\QC^{(1)}$ satisfies an inequality,
\begin{equation}
    \QC^{(1)}(\BC_1 \ot \BC_2) \geq \QC^{(1)}(\BC_1) + \QC^{(1)}(\BC_2),
    \label{nonAddEq}
\end{equation}
which can be strict~\cite{DiVincenzoShorEA98}. Like $C^{(1)}$, $\QC^{(1)}$ is
an entropic quantity, however it represents an achievable rate for correcting
errors in quantum information sent across a noisy quantum channel. A channel's
quantum capacity $\QC$ is defined to be the best possible achievable
rate~\cite{BennettDiVincenzoEA96}. Non-additivity of $\QC^{(1)}$ makes $\QC$
difficult to compute~\cite{WolfCubittEA11, OskoueiMancini18}, and more
markedly it makes $\QC$ an incomplete measure~\cite{SmithYard08} of a channel's
ability to send quantum information.

The difficulty in computing $\QC$ essentially comes from a
strict inequality in~\eqref{nonAddEq}, found when $\BC_1$ and $\BC_2$ are tensor
products of the same channel $\BC$. Low dimensional channels which display this
type of non-additivity include a variety of very noisy qubit channels including
the depolarizing~\cite{DiVincenzoShorEA98, FernWhaley08}, the
dephrasure~\cite{LeditzkyLeungEA18} and other qubit
Pauli~\cite{SmithSmolin07,FernWhaley08, BauschLeditzky19} and generalized
erasure channels~\cite{SiddhuGriffiths20, Filippov21}. As a result, even when a channel
$\BC$ is relatively simple, its quantum capacity $\QC(\BC)$ must be obtained as
the limit $n \mapsto \infty$ of a sequence $\QC^{(1)}(\BC^{\ot
n})/n$~\cite{Lloyd97, BarnumKnillEA00, Shor02a, CaiWinterEA04, Devetak05}. This
limit, sometimes called a {\em regularization of $\QC^{(1)}$}, can be
particularly intractable: there are very noisy high dimensional channels for
which each term in this sequence can be larger than the previous
one~\cite{ElkoussStrelchuk15}.  In addition, for any integer $k$ there is a
channel $\tilde \BC$ for which $\QC^{(1)}(\tilde \BC^{\ot k}) = 0$ but
$\QC(\tilde \BC)>0$~\cite{CubittElkoussEA15}. This type of unbounded
non-additivity makes it hard to even check if a channel's quantum capacity is
strictly positive or zero.

Challenges in computing and checking positivity of a channel's quantum capacity
can be circumvented in the special case of
(anti)-degradable channels~\cite{DevetakShor05, CubittRuskaiEA08}, PPT
channels~\cite{HorodeckiHorodeckiEA00}, DSPT channels~\cite{GaoJungeEA18a},
and less noisy channels~\cite{Watanabe12}. However, even if one computes a
channel's quantum capacity, non-additivity implies that this capacity
may be an incomplete measure of the channel's ability to send quantum
information. Instances of non-additivity, i.e., a strict inequality
in~\eqref{nonAddEq}, have been found when $\BC_1$ and $\BC_2$ are different
channels, each having no quantum capacity. One instance, called {\em
superactivation} has been found when $\BC_1$ is a PPT channel and $\BC_2$ is a
zero capacity erasure or depolarizing channel~\cite{SmithYard08,
BrandaoOppenheimEA12}.  Another instance of non-additivity has been found where
$\BC_1$ is a rocket channel and $\BC_2$ is an erasure
channel~\cite{SmithSmolin09a}, both channels are again very noisy, $\BC_1$ has
small quantum capacity while $\BC_2$ has none, but together they have coherent
information much larger than the sum of quantum capacities of each channel. 

In the past, instances of non-additivity found in very noisy
channels have shown that quantum information and channels can display a type of
synergy which is absent from their classical counterparts.  
Non-additvity has previously not been found in low-noise channels, those with
coherent information comparable to the quantum capacity of a perfect~(identity)
channel with the same input dimension as the channel.  By contrast, in certain
low-noise channels non-additivity has been shown to be
absent~\cite{BrandaoEisertEA11}, and in low-noise Pauli channels non-additivity
has been shown to be of little practical relavance~\cite{LeditzkyLeungEA18a}.
While the study of non-additivity remains of fundamental interest, methods for
finding and exploring non-additivity are scant.  In high
dimensional and high noise PPT and rocket channels, non-additivity is
found by using the special structure of these channels. Whereas
in qubit and other low dimensional but high noise channels, methods based
on degenerate quantum codes~\cite{SmithSmolin07} and numerical
searches~\cite{BauschLeditzky20} can identify non-additivity, but these too can
falter in simple cases of interest~\cite{SiddhuGriffiths20}. 

Strategies to check if a general channel has zero or non-zero quantum capacity
are limited~\cite{SmithSmolin12, StrelchukOppenheim12}.  To test if a channel
has zero quantum capacity, one can check whether the channel is PPT or
anti-degradable.  For special channels these two checks can be done
algebraically~\cite{WolfPerezGarcia07, CubittRuskaiEA08, StrelchukOppenheim12},
but in general, they require numerically solving a semi-definite
program~\cite{SutterScholzEA17}.  Even if one performs these checks, their
results can be inconclusive because there may exist zero capacity channels that
are neither PPT nor anti-degradable.  Testing if a channel has non-zero quantum
capacity is tricky.  Except in very special circumstances, there are no
algebraic tests.  Numerics can be used to check if a channel's coherent
information is non-zero. However, these numerics can be unreliable, even for
low dimensional qubit channels~\cite{LeungWatrous17, Wilde18,
LeditzkyLeungEA18} without unbounded non-additivity. For high dimensional
channel's numerics can be expensive~\cite{Fern08,FernWhaley08} in addition to
being unreliable~\cite{CubittElkoussEA15}.

Seeking physical and mathematical mechanisms to find and understand positivity
and non-additivity remains an enduring challenge in quantum information
science. While this challenge tempers hopes for rapid progress on understanding
quantum capacities, it also presents an opportunity to introduce new ideas for
addressing this challenge.

In this work, we introduce a simple but key property of the von-Neumann
entropy, we call it a log-singularity~(see Fig.~\ref{Fig1LS}), and show that
changes in entropy caused by log-singularities are a mathematical mechanism
responsible for both positivity and non-additivity of the coherent information.
Utilizing this mechanism, (1) we provide an instance of non-additivity using a
zero capacity qubit channel in parallel with a low-noise qutrit channel with
$\QC^{(1)}/\log_2 3 \simeq .6$;
(2) we prove a general theorem which gives algebraic conditions under which a
quantum channel must have strictly positive coherent information. A corollary
of this theorem gives a simple, dimensional test for capacity. An application
of this corollary provides a characterization of all qubit channels whose
complement has non-zero quantum capacity. A separate application of the theorem
reveals how a large class of zero capacity qubit channels can assist an
incomplete erasure channel in sending quantum information.

\section{Results}
\label{SRes}

\begin{figure}[ht]
    \centering
    \includegraphics[scale=.88]{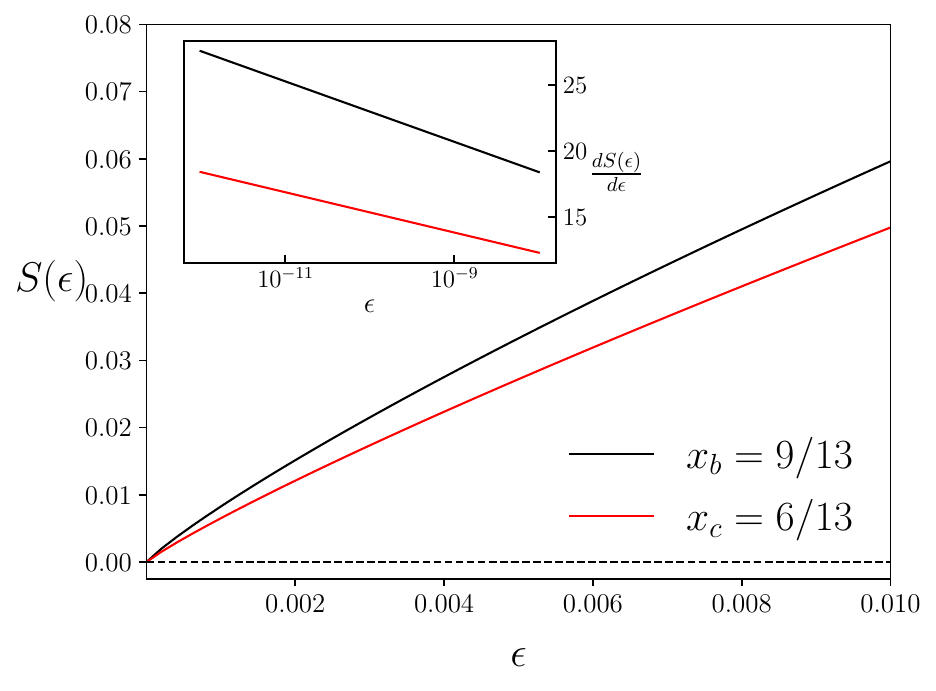}
    \caption{{\bf Behaviour of the von-Neumann entropy in the vicinity of an
    $\ep \log$-singularity.} Two density operators $\rho_b(\ep)$ and $\rho_c(\ep)$
    with spectrum $(1-x_b \ep, x_b \ep)$ and $(1-x_c \ep, x_c \ep/3, x_c \ep/3, x_c
    \ep/3)$ respectively have entropies $S_b(\ep)$ and $S_c(\ep)$ respectively,
    where $0 \leq \ep \leq 1$. For fixed $\ep \log$-singularity rates
    $x_b = 9/13$ and $x_c = 6/13$, of $S_b(\ep)$ and $S_c(\ep)$, respectively, a
    plot of these entropies $S(\ep)$ as a function of $\ep$. The inset shows
    the gradient of the entropies as a function of $\log_2 \ep$ for small $\ep$.}
    \label{Fig1LS}
\end{figure}

{\bf Log-singularity}.---Let $\rho(\ep)$ denote a density operator that depends
on a real positive parameter $\ep$, and $S(\ep) = -\Tr\big( \rho(\ep) \log
\rho(\ep) \big)$ denote its von-Neumann entropy.  If one or several
eigenvalues of $\rho(\ep)$ increase linearly from zero to leading order in
$\ep$ then a small increase in $\ep$ from zero increase $S(\ep)$ by $x|\ep
\log\ep|$ for some constant $x>0$; i.e., $dS(\ep)/d \ep \simeq -x \log \ep$,
and we say $S(\ep)$ has an {\em $\ep \log$-singularity} with {\em rate $x$}.
For instance a qubit density operator with spectrum $(1-x_b \ep, x_b \ep),
0 \leq \ep \leq 1$ and $0 \leq x_b \leq 1$ has an $\ep \log$-singularity of
rate $x_b$.  A quqart density operator with spectrum $(1-x_c \ep, x_c \ep/3,
x_c \ep /3, x_c \ep /3)$ and $0 \leq x_c \leq 1$, has an $\ep \log$-singularity
of rate $x_c$~(also see Fig.~\ref{Fig1LS}).

The term $\ep \log$-singularity comes from the behaviour where the
derivative of $S(\ep)$ with respect to $\ep$ is logarithmic in $\ep$ and this
derivative tends to infinity as $\ep$ tends to zero. While $S(\ep)$ is
continuous, behaviour of continuity bounds on $S(\ep)$ can be dominated by $\ep
\log$-singularities in the sense that changes in continuity bounds can be
essentially logarithmically in $\ep$ for small $\ep$~(see Appendix~\ref{ACnt}).
For very small $\ep$, this singularity causes a sharp change in the von-Neumann
entropy. Since this sharp change occurs for very small parameter values $\ep$,
its effects can be prohibitively hard to detect numerically. However, when
these effects appear, they dominate the behaviour of the von-Neumann entropy,
and the physics which may directly depend on this entropy.

Understanding of the physics of sending noiseless quantum information across a
noisy quantum channel is aided by the channel's coherent information
$\QC^{(1)}$. 
To define a quantum channel and its coherent information, consider an
isometry $J: a \mapsto b \ot c$ that generates a pair of quantum channels
$\BC:a \mapsto b$ and $\CC : a \mapsto c$, where each channel may be called the
complement of the other.  These channels map an input density operator $\rho_a$
to outputs $\rho_b:=\BC(\rho_a) = \Tr_{c}(J \rho_a J^\dag)$ and
$\rho_c:=\CC(\rho_a) = \Tr_{b}(J \rho_a J^\dag)$, respectively. 
The dimensions $d_b$ and $d_c$, of outputs $b$ and $c$, respectively are the
ranks of $\BC(I_a)$ and $\CC(I_a)$, respectively. These are the smallest
possible output dimensions required to define $\BC$ and $\CC$~(in the notation
of Def.~4.4.4 in~\cite{Wilde17}, $d_b$ is the Choi-rank of $\CC$ and $d_c$ is
the Choi-rank of $\BC$). These definitions make the channel pair setting
symmetric with respect to replacement of one channel in the pair with its
complement.
The coherent information~(or the entropy bias) of $\BC$ at $\rho_a$, $\Dl(\BC,
\rho_a):= S(\rho_b) - S(\rho_c)$, maximized over density operators $\rho_a$
gives the {\em channel coherent information} $\QC^{(1)}(\BC)$.

\begin{figure}[ht]
    \centering
    \includegraphics[scale=1]{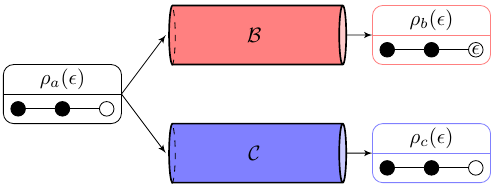}
    \caption{{\bf Schematic for the reason behind an $\ep\log$-singularity.} An
    input density operator $\rho_a(\ep)$ is mapped by a channel $\BC$ to a
    density operator $\rho_b(\ep)$ and by the channel's complement $\CC$ to
    a different density operator $\rho_c(\ep)$. Below each operator is a
    representation of its spectrum where closed and open circles indicate
    eigenvalues that, for all $0 \leq \ep \leq 1$, are non-zero and zero
    respectively. A circle with $\ep$ indicates an eigenvalue that increases
    linearly from zero to leading order in $\ep$. Since the spectrum of
    $\rho_b(\ep)$ has a circle with $\ep$, its von-Neumann entropy has an
    $\ep \log$-singularity.}
    \label{FigA}
\end{figure}

When considering an input density operator, $\rho_a(\ep)$, we use a concise
notation $S_b(\ep):= S\big(\rho_b(\ep)\big)$, $S_c(\ep) :=
S\big(\rho_c(\ep)\big)$ and $\Dl (\ep) := \Dl\big(\BC, \rho_a(\ep)\big) =
S_b(\ep) - S_c(\ep)$.
At $\ep = 0$ if $\rho_a(\ep)$ has rank $d_a$, then by definition of $d_b$, rank
of $\rho_b(0)$ will be $d_b$~(see Appendix.~\ref{AppAR1}), as a
result $S_b(\ep)$ will not have an $\ep \log$-singularity.  A similar argument
shows if $\rho_a(0)$ is rank $d_a$ then $S_c(\ep)$ does not have an $\ep
\log$-singularity. 
When rank of $\rho_a(0)$ is strictly less than $d_a$, then an $\ep
\log$-singularity can be present in $S_b(\ep)$ or $S_c(\ep)$~(see
Fig.~\ref{FigA}), or an $\ep \log$-singularity can be present in both
$S_b(\ep)$ and $S_c(\ep)$ in which case the $\ep \log$-singularity with larger
rate is said to be {\em stronger}.

\begin{figure} 
    \centering
    \includegraphics[scale=.7]{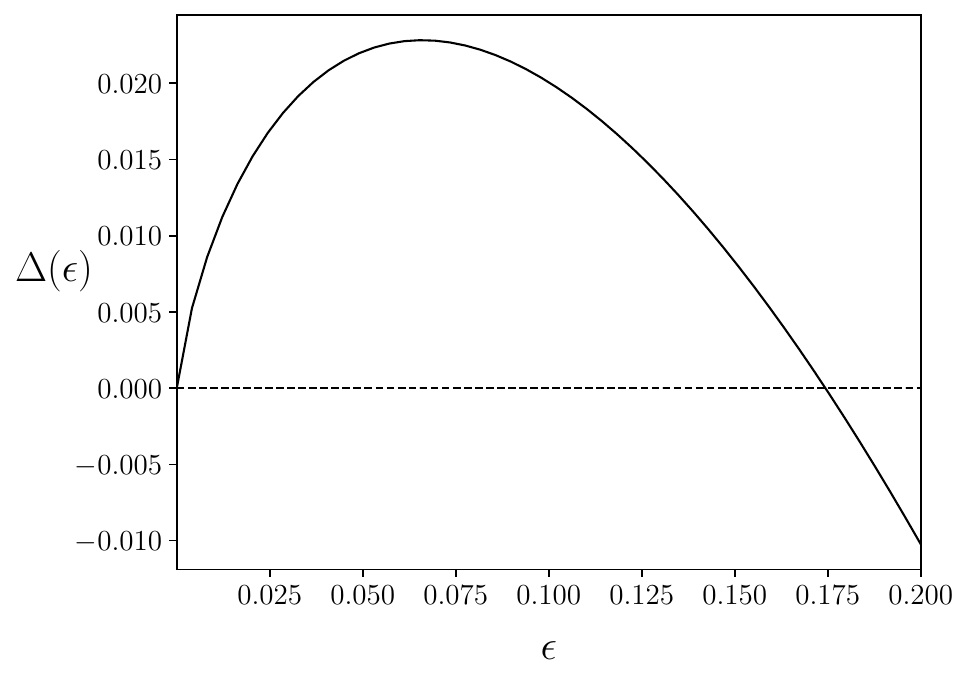}
    \caption{{\bf Illustration of $\log$-singularity based mechanism behind
    positivity of the channel coherent information.} For the channel $\BC$ defined
    by isometry in eq.~\eqref{LIso}, the entropy difference $\Dl(\ep) = S_b(\ep) -
    S_c(\ep)$ for density operators below eq.~\eqref{LIso} as a function of $\ep$
    is plotted above.  Here $S_b(\ep)$ and $S_c(\ep)$ have the same entropy at $\ep
    = 0$ and they both have $\ep \log$-singularities. The singularity in $S_b(\ep)$
    has a rate $x_b = 9/13$ which is higher than $x_c = 6/13$, the rate of the
    singularity in $S_c(\ep)$. This higher rate $x_b$ makes both $\Dl(\ep)$ and
    $\QC^{(1)}(\BC)$ strictly positive for small $\ep$, even though for larger
    $\ep$, $\Dl(\ep)<0$.}
\label{FigLSDiff}
\end{figure}

{\bf Positivity of Coherent Information}.--- Changes in the von-Neumann entropy
caused by $\log$-singularities can act as a mechanism which makes
$\QC^{(1)}(\BC)>0$~(see Fig.~\ref{FigLSDiff}).
To illustrate this mechanism consider a convex combination of input density
operators $\hat \rho_a$ and $\sg_a$:
\begin{equation}
    \rho_a(\ep) = (1-\ep) \hat \rho_a + \ep \sg_a, \quad \ep \in [0,1].
    \label{rhoInDef}
\end{equation}
This convex combination~\eqref{rhoInDef} leads to other such combinations,
\begin{equation}
    \rho_b(\ep) = (1-\ep)\hat \rho_b + \ep \sg_b \quad \text{and} 
    \quad
    \rho_c(\ep) = (1-\ep)\hat \rho_c + \ep \sg_c,
    \label{rhoOutDef}
\end{equation}
at the outputs of $\BC$ and $\CC$, respectively.  
Let $\hat \rho_a$ be a pure state, then $\Dl(\BC, \hat \rho_a) = 0$ i.e.,
$\Dl(0) = 0$~(see Appendix~\ref{AppAR1}). Assume $\hat \rho_a,
\sg_a,$ and the channel pair $(\BC, \CC)$ are such that an $\ep
\log$-singularity is present in $S_b(\ep)$ but not in $S_c(\ep)$; that is, for
a small enough increase in $\ep$ from zero, $S_b(\ep)$ increases by $|O(\ep
\log \ep)|$ but $S_c(\ep)$ has no $O(\ep \log \ep)$ increase. Thus, for small enough
$\ep$, $\Dl(\ep) \simeq |O(\ep \log \ep)|>0$; since $\Dl(\ep) \leq
\QC^{(1)}(\BC)$, we conclude $\QC^{(1)}(\BC) > 0$.

In the illustration above, let the channels $\BC:a \mapsto b$ and $\CC : a
\mapsto c$ be defined by an isometry $L:a \mapsto b \ot c$ of the form,
\begin{equation}
    L \ket{0} = \ket{00}, \quad
    L \ket{1} = \sqrt{\frac{2}{9}} \ket{01} + \sqrt{\frac{7}{9}} \ket{10},
    \quad \text{and} \quad
    L \ket{2} = \frac{1}{\sqrt{2}}(\ket{02} + \ket{13}),
    \label{LIso}
\end{equation}
where $\{ \ket{i} \}$ represents the standard basis, and $\ket{ij}$ denotes
$\ket{i} \ot \ket{j} \in b \ot c$.  Let $[\psi]$ denote the dyad $\dya{\psi}$.
A channel input of the form in~\eqref{rhoInDef} with $\hat \rho_a = [0]$ and
$\sg = (9[1] + 4[2])/13$ leads to channel outputs of the
form~\eqref{rhoOutDef}.  These outputs $\rho_b(\ep)$ and $\rho_c(\ep)$ have
$\ep \log$-singularities with rates $x_b = 9/13$ and $x_c = 6/13$,
respectively~(see Fig.~\ref{Fig1LS}). Since $x_b > x_c$, $\QC^{(1)}(\BC)>0$, as
shown in Fig~\ref{FigLSDiff}.

In general, a $\log$-singularity based mechanism can make $\QC^{(1)}(\BC) > 0$
if there is a channel input $\rho_{a}(\ep)$ for which $\Dl(0) = 0$, and either
there is an $\ep\log$-singularity in $S_b(\ep)$ but not in $S_c(\ep)$ or there
is an $\ep\log$-singularity in both $S_b(\ep)$ and $S_c(\ep)$ but the one in
$S_b(\ep)$ is stronger, in either case, an argument similar to the one in our
illustration above implies that $\QC^{(1)}(\BC)>0$. An analogous
$\log$-singularity based mechanism can make $\QC^{(1)}(\CC) > 0$.
In principle this mechanism can be applied to a quantum channel $\BC$,
regardless of how small or large $\QC^{(1)}(\BC)$ may be.  In practice, we find
that the above mechanism applies in a general situation presented next.

\begin{theorem}
\label{thrm}
If a quantum channel $\BC$, with output and environment dimension $d_b$ and
$d_c$ respectively, maps some pure state to an output of rank $d_c<d_b$, 
then $\QC^{(1)}(\BC) > 0$.
\end{theorem}
This theorem applies quite generally, including cases where $d_b \geq d_c$. In
these $d_b \geq d_c$ cases, a channel $\BC$'s coherent information is strictly
positive if the theorem holds for any sub-channel of $\BC$. There are simple
examples~(for instance see Appendix.~\ref{AThrmAps}) of
channels with $d_b = d_c$ where the above theorem applies.

\begin{figure}[ht]
    \centering
    \includegraphics[scale=1]{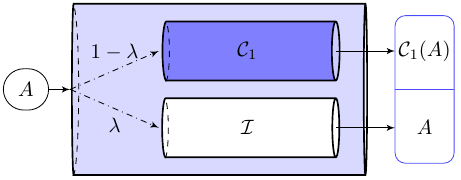}
    \caption{{\bf Incomplete erasure channel $\CC$.} Channel $\CC$'s input $A$,
    with probability $\lm$ goes via $\IC$~(the identity channel) to an output
    subspace as $A$, or else via $\CC_1$~(a noisy channel) to an orthogonal
    subspace as $\CC_1(A)$.
    When $\CC_1 = \TC$~(where $\TC(A) = \Tr(A) \dya{0}$) then $\CC$ becomes the
    usual erasure channel with erasure probability $1-\lm$, whose quantum
    capacity is zero for $\lm \leq 1/2$.
    However, an application of Theorem~\ref{thrm} shows that as
    $\CC_1$ is changed from $\TC$ to one of several different zero capacity qubit
    channels, coherent information of $\CC$ becomes positive for all $\lm>0$.}
    \label{FigB}
\end{figure}

Theorem~\ref{thrm} can be applied to the incomplete erasure channel $\CC(\rho) =
\lm \rho \oplus (1- \lm) \CC_1(\rho)$~\cite{SiddhuGriffiths20} whose output is
split into two orthogonal subspaces~(see Fig.~\ref{FigB}). The channel's input
is sent unchanged to the first subspace with probability $\lm$, else it is sent
via a noisy channel $\CC_1$ to the second subspace.  This channel $\CC$ is
relevant for describing noise in experiments where the channel user knows if
noise has acted or not. 

Suppose the incomplete erasure channel $\CC$ has a qubit input and $\CC_1$ is
any zero quantum capacity qubit channel with a qubit
environment~\cite{WolfPerezGarcia07}. Any such qubit channel $\CC_1$ has a
noise parameter $0 \leq p \leq 1/2$; where at $p=0$, $\CC_1$ erases its input
by taking it to a fixed pure state, making $\CC$ a regular erasure channel.
This regular erasure channel has zero coherent information, i.e.,
$\QC^{(1)}(\CC) = 0$ when $\lm$ is below a threshold $\lm_0 =
1/2$~\cite{BennettDiVincenzoEA97}. As noise in $\CC_1$ is decreased by
continuously increasing $p$, this threshold is expected to decrease
continuously.  While ordinary numerics may seem to confirm this expectation, a
simple use of Theorem~\ref{thrm} shows that for any $p>0$, $\QC^{(1)}(\CC) > 0$ for
any $\lm>0$, i.e., an arbitrarily small increase in $p$ from zero shifts the
threshold value from $\lm_0 = 1/2$ to $\lm_0 = 0$.  This discontinuous shift
doesn't appear in standard numerics because for small $p$, $\QC^{(1)}(\CC)$ can
be as small as $O(e^{-1000})$~\cite{SiddhuGriffiths20}, a number much beyond
ordinary numerical precision. Such discontinuous shifts reveal an unexpected
behaviour: a channel $\CC_1$ which can't send quantum information on its own,
i.e, it has no quantum capacity, can nonetheless assist an incomplete erasure
channel in sending quantum information. Such assistance was found previously
for two specific qubit channels $\CC_1$ using arguments tailored for those
specific channels~\cite{SiddhuGriffiths20, LeditzkyLeungEA18}.  Our argument
here generalizes those results and points to $\log$-singularities as a generic
mathematical cause behind this assistance.  This assistance is particularly
intriguing because it occurs for a wide variety of zero capacity qubit channels
$\CC_1$ but doesn't occur for arbitrary zero capacity channels. For instance
when $\CC_1$ is a zero quantum capacity erasure channel with arbitrary input
dimension and erasure probability $\mu \geq 1/2$, $\QC^{(1)}(\CC) = 0 $ for $0 \leq \lm
\leq 1- 1/(2 \mu)$.

To check if a channel $\BC:a \mapsto b$ has strictly positive coherent
information one may numerically find an input density operator $\rho$ for which
the entropy difference $\Dl(\BC,\rho)>0$. This numerical search can be
unreliable because $\Dl(\BC,\rho)$ is generally non-convex in $\rho$ and
$\Dl(\BC, \rho)$ can be affected by $\log$-singularities. The search can also
be expensive for channels with large input and output dimensions.  A corollary
of Theorem~\ref{thrm} gives an algebraic result
revealing a wide and simple set of channels with
large output dimension and non-zero coherent information:
\begin{corollary}
\label{cr1}
Any channel $\BC$ has $\QC^{(1)}(\BC)>0$ if its input dimension $d_a > 1$ and
output dimension $d_b > d_a(d_c - 1)$.
\end{corollary}
This result is easy to apply: given a channel one simply
uses its dimensions to check if the channel satisfies the conditions of
Corollary~\ref{cr1}.  For instance, 
Corollary~\ref{cr1} implies that any channel whose output dimension is
larger than its input dimension has strictly positive $\QC^{(1)}$ whenever the
channel's environment is a qubit, i.e., $d_c=2$.

Qubit channels are extremely useful for characterizing noise in experiments.
Despite the vast body of work dedicated to studying the capacity of qubit
channels~\cite{RuskaiSzarekEA02, KingRuskai01, King02, King03,
KhatriSharmaEA20, WolfPerezGarcia07}, a basic question has remained open: when
does the complement of a qubit channel have non-zero quantum capacity? 
Corollary~\ref{cr1}, in conjunction with prior work~\cite{WolfPerezGarcia07}, answers
this question.  If the complement of a qubit channel has output
dimension 1 or 2, then conditions under which this complement has non-zero
capacity can be found in~\cite{WolfPerezGarcia07}; for all remaining cases, 
Corollary~\ref{cr1}~(see Appendix~\ref{AThrmAps}) shows that
complement has strictly positive coherent information and quantum capacity.
This positivity result contains as a special case the results
of~\cite{LeungWatrous17} which showed that any qubit Pauli channel $\BC$ with
$d_c= 3$ or $4$ has a complement with positive channel coherent information.

{\bf Non-additivity of Coherent Information}: A mechanism based on
$\log$-singularities can give rise to non-additivity of $\QC^{(1)}$. For
two~(possibly different) quantum channels $\BC_1$ and $\BC_2$, let
\begin{equation}
    \QC^{(1)}(\BC_1) = \Dl(\BC_1, \rho_{a1}^*), \quad \text{and} \quad
    \QC^{(1)}(\BC_2) = \Dl(\BC_2, \rho_{a2}^*),
    \label{nonAdd1}
\end{equation}
for some density operators $\rho_{a1}^*$ and $\rho_{a2}^*$ and let $\BC :=\BC_1
\ot \BC_2$. Choose $\rho_{a}(\ep)$ at the input of $\BC$ with the property that $\rho_a(0)
=\rho_{a1}^* \ot \rho_{a2}^*$ and $S_b(\ep)$ has a stronger $\ep
\log$-singularity than $S_c(\ep)$, then a small enough increase in $\ep$ from zero will
increase $\Dl(\ep)$ from $\QC^{(1)}(\BC_1) + \QC^{(1)}(\BC_2)$ by $|O(\ep \log
\ep)|$ indicating a strict inequality in~\eqref{nonAddEq}.
This $\log$-singularity based mathematical mechanism responsible for
non-additivity requires $S_b(\ep)$ to have an $\ep \log$-singularity. As stated
earlier, this requirement can be satisfied if $\rho_a(0)$ has less than full
rank. A condition satisfied by several channels with zero and non-zero coherent
information.  This mechanism will now be used in an explicit instance of
non-additivity using two channels, one with zero and another with large
positive coherent information.

To present this instance of non-additivity, we introduce a low-noise qutrit
channel $\BC_1$ whose coherent information is comparable to that of a qutrit
identity channel.
This channel's superoperator $\BC_1(\rho) = \Tr_{c1}(J_1 \rho J_1^{\dag})$
comes from an isometry $J_1: a1 \mapsto b1 \ot c1$ of the form,
\begin{equation}
    J_1 \ket{0} = \sqrt{s} \ket{00} + \sqrt{1-s} \ket{11},
    \quad
    J_1 \ket{1} = \ket{21},
    \quad \text{and} \quad
    J_1 \ket{2} = \ket{20},
    \label{qTritIso}
\end{equation}
where $0 \leq s \leq 1$.  Since an exchange of $s$ with $1-s$ can be achieved
by local unitaries in $a1,b1$ and $c1$, we restrict ourselves to $0 \leq s \leq
1/2$. The channel coherent information $\QC^{(1)}(\BC_1)$ is given by its
entropy difference $\Dl(\BC_1, \rho_{a1}^*)$ where $\rho_{a1}^* = (1-w)[0] + w
[1]$, $0<w<1$~(see Appendix~\ref{AQCI}).  At $s = 0$, $\QC^{(1)}(\BC_1) = 1$
and decreases monotonically with the noise parameter $s$ to become $\simeq
.695$ at $s = 1/2$.  These values of $\QC^{(1)}(\BC_1)$ bound the quantum
capacity of $\BC_1$ from below and they are comparable to the quantum capacity,
$\log_2 3$, of the qutrit identity channel.

\begin{figure}[ht]
    \centering
    \includegraphics[scale=.7]{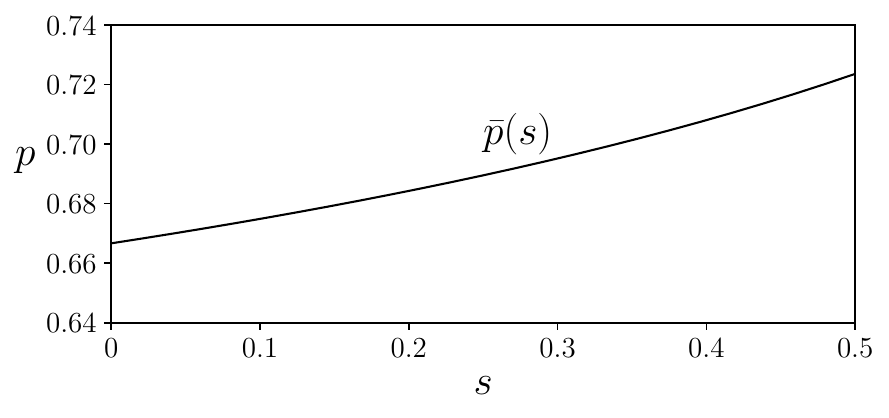}
    \caption{{\bf Non-additivity in a low-noise channel.} For the low-noise
    qutrit channel $\BC_1$ with noise parameter $0 < s \leq 1/2$, $.695 \leq
    \QC^{(1)}(\BC_1) < 1$ and qubit amplitude damping channel $\BC_2$ with damping
    probability $1/2 \leq p \leq 1$, $\QC^{(1)}(\BC_2) = \QC(\BC_2) = 0$, we find
    $\QC^{(1)}(\BC_1 \ot \BC_2)$ to be strictly larger than $\QC^{(1)}(\BC_1) +
    \QC^{(1)}(\BC_2)$ when $ 1/2 \leq p \leq  \bar p(s)$ with $\bar p(s)$ plotted
    above.  \label{FigD}}
\end{figure}

A $\log$-singularity based argument stated earlier shows that using $\BC_1$ in
parallel with $\BC_2$, a zero quantum capacity qubit amplitude damping channel
with damping probability $p \geq 1/2$, results in non-additivity, i.e., a
strict inequality in~\eqref{nonAddEq} for all $0 < s \leq 1/2$ and $1/2 \leq p
< \bar{p}(s)$~(see Fig.~\ref{FigD} and Sec.~\ref{SMeth}). 
This non-additivity has several interesting features. 
First, it shows existence of previously unknown non-additivity when using
a low-noise channel.
Second, the non-additivity reported here fosters a more nuanced understanding
of quantum capacity. Even in a setting using very simple low-dimensional
channels, the quantum capacity provides an incomplete description of a
channel's ability to send quantum information. As shown here, despite having no
quantum capacity, the qubit amplitude damping channel $\BC_2$ does posses a
separate ability to assist transmission of quantum information when used in parallel with
a simple qutrit channel $\BC_1$.
Third, the non-additivity here is robust against amplitude damping noise:
additional amplitude damping noise beyond $p = 1/2$ does not immediately
destroy this non-additive effect which survives till $p < \bar{p}(s)$. Fourth,
this instance of non-additivity has a very wide range: it is present over the
entire parameter space of the qutrit channel $\BC_1$, except at a single point
$s = 0$. Fifth, numerical techniques, which are commonly used to find
non-additivity, can easily miss this wide range of non-additivity which appears
because of changes in entropy caused by $\log$-singularities.

\section{Discussion}

We have discussed logarithmic~($\log$) singularities that occur quite generally
in the von-Neumann entropy of any density operator being moved linearly from
the boundary to the interior of the set of density operators. In the region
where this singularity occurs, it dominates the behaviour of the von-Neumann
entropy.
This kind of dominance can be used to extract insights about the physics which
depends on this entropy. We have investigated the physics of sending quantum
information.  Our investigation leads to an insight that
$\log$-singularities act as a mathematical source behind both positivity and
non-additivity in a channel's coherent information.  An analysis of
$\log$-singularities could potentially be useful in other areas where the
von-Neumann entropy plays a central role. 
One area of this type is the study of continuous variable channels.
Capacities of these channels remain an active area of
research~\cite{HolevoWerner01, GiovannettiLloydEA03, BraunsteinLoock05,
CarusoGiovannetti06, Holevo07, WolfPerezGarciaEA07, PirandolaLaurenzaEA17}, and
these capacities also display a variety of exotic
behaviour~\cite{NohPirandolaEA20}, including
superactivation~\cite{SmithSmolinEA11, LimLee18, LimTakagiEA19}.  Extending our
$\log$-singularity ideas to investigate such exotic behaviour would be an
interesting direction of future work.

Checking if any general channel has non-zero or zero quantum capacity is a
fundamental but hard problem. While some general methods have been proposed to
solve this problem~\cite{SmithSmolin12}, they are not always easy to apply and
don't necessarily lead to new channels with zero or positive quantum capacity. 
By contrast the algebraic $\log$-singularity based method proposed here is easy
to apply and it unearths a variety of channels with positive quantum
capacity. Using it, we give Theorem~\ref{thrm} which
reveals certain general conditions for strict positivity of a channel's
coherent information. Corollary~\ref{cr1} of Theorem~\ref{thrm}
unearths a wide variety of channels with strictly positive coherent
information. Corollary~\ref{cr1} only makes use of a
channel’s dimensions. Extending this corollary, for example by showing
$\QC^{(1)}(\BC) > 0$ for some $d_b > d_c$, would be an interesting direction of
future research.
Yet another direction would be to supplement our mathematical
$\log$-singularity reasoning with more physical arguments. 
Such reasoning may help explain the positivity of $\QC^{(1)}$ found
in the incomplete erasure channel and clarify why the simplest zero capacity
channels behave differently from others when used as part of the incomplete
erasure channel. This clarification may provide further insights into the
transmission of quantum information.
Another source of insight may be a quantitative analysis of bounds on the
quantum capacity~\cite{LeungSmith09, Ouyang14, SutterScholzEA17, CrossLiEA17,
GaoJungeEA18, LeditzkyDattaEA18, LeditzkyLeungEA18a, FanizzaKianvashEA20} of
the incomplete erasure channel or some other channel where a $\log$-singularity
based mechanism is responsible for strict positivity of the channel's coherent
information. In certain cases, $\log$-singularities can dominate continuity
based bounds on the coherent information and it would be interesting to see if
such effects also appear in continuity bounds on a channel's quantum
capacity~\cite{LeungSmith09}.

Using $\log$-singularities, we have shown that the coherent information and
quantum capacity of several channels is non-zero. It follows that the two-way
quantum capacity~\cite{BennettDiVincenzoEA96} and the private
capacity~\cite{Devetak05} of these same channels are also non-zero.  
These observations comes from the simple fact that the quantum capacity of a
channel is a lower bound on the channel's private and two-way quantum
capacities.
These other capacities are even less understood than the quantum capacity and
our $\log$-singularity based analysis could prove useful in their
investigation. For instance, the reverse coherent
information~\cite{GarciaPatronPirandolaEA09}, which is a lower bound on the
two-way quantum capacity, may yield to a $\log$-singularity based analysis,
similar to the one performed here. 
Admittedly, our $\log$-singularity based method for showing positivity of
capacity does not solve the general problem of finding all channels with
strictly positive capacity. Results concerning unbounded
non-additivity~\cite{CubittElkoussEA15, ElkoussStrelchuk15} temper hopes about
the existence of an easy to apply but completely general method for checking
positivity of the quantum capacity.  Our work nonetheless points out that such
tempering need not hinder progress in finding new, interesting, and potentially
insightful instances of channels with strictly positive quantum capacity.

Another aspect of our findings is how changes in the von-Neumann entropy
casued by $\log$ singularities is a mathematical mechanism responsible for
non-additivity. Prior search for such mechanisms have focussed on the structure
of special channels~\cite{SmithYard08, BrandaoOppenheimEA12, SmithSmolin09a}
used for obtaining non-additivity or on the use of certain tailored quantum
codes~\cite{SmithSmolin07, SmithRenesEA08,BauschLeditzky20}.  We open another
direction by showing how a fundamental property of the von-Neumann entropy can
lead to non-additivity. This $\log$-singularity property can be analyzed to
find non-additivity in channels with large, small or no quantum capacity. The
algebraic nature of this analysis allows us to identify non-additivity over
wide ranges of a channel's parameter, without the need for traditional
numerics~\cite{Fern08, FernWhaley08, BauschLeditzky19}. While our work opens
one path, it is not the only path forward. We leave open the exciting but
challenging possibility of finding other mathematical and physical principles
that may explain non-additive effects in quantum information science. 

Unlike prior work, non-additivity of the coherent information reported here
occurs in a low-noise channel.
From a fundamental physics perspecitve, non-additivity using the product of one
low-noise channel with another low-dimensional but zero capacity qubit
amplitude damping channel is surprising because it implies that even in such a
simple setting the quantum capacity of the amplitue damping channel does not
fully characterize its resourcefullness for sending quantum information.

This simple example adds to the collection of exotic channels from which
further physics can be extracted. In principle, this low-dimensional and
low-noise channel can be experimentally realized.  Given the practical
relevance of low-noise channels, our finding of non-additivity in such channels
suggests that non-additivity is not just a fundamental curiosity but a
potential resource for quantum technologies. 

\section{Methods}
\label{SMeth}

{\bf Proof of Theorem~\ref{thrm}}:
A $\log$-singularity based mechanism responsible for making a channel's
coherent information strictly positive is the key ingredient in the proof of
Theorem~\ref{thrm}.
Assume $d_c<d_b$ and $\BC$ maps some pure state $[\psi]_a$ to an output
$\BC([\psi]_a)$ of rank $d_c$.
Consider an input density operator $\rho_a(\ep)$ of the form in
eq.~\eqref{rhoInDef} where $\hat \rho_a$ is the pure state $[\psi]_a$ and
$\sg_a = I_a/d_a$. The outputs $\rho_b(\ep)$ and $\rho_c(\ep)$ have the
form in~\eqref{rhoOutDef} where $\hat \rho_b$ and $\hat \rho_c$ have the
same rank $d_c$, $\sg_c$ and $\sg_b$ have ranks $d_c$
and $d_b$ respectively~(see para~3 in Sec.~\ref{SRes}). As a result
$S_b(\ep)$ has an $\ep \log$-singularity while $S_c(\ep)$ doesn't,
consequently $\QC^{(1)}(\BC)>0$.
The absence of an $\ep \log$-singularity in $S_c(\ep)$ follows from the fact
that at $\ep = 0$, $\rho_c(\ep)$ is a rank $d_c$ operator $\hat \rho_c$. To
notice the presence of an $\ep \log$-singularity in $S_b(\ep)$, it is helpful
to rewrite $\rho_b(\ep)$ in eq.~\eqref{rhoOutDef}, as
\begin{equation}
    \rho_b(\ep) = \hat \rho_b + \ep \om_b,
\end{equation}
where 
\begin{equation}
    \om_b := \sg_b - \hat \rho_b.
\end{equation}
At $\ep = 0$, $\rho_b(\ep)$ is $\hat \rho_b$, which has $d_b - d_c$ zero
eigenvalues. Corresponding to these zero eigenvalues is an eigenspace of
dimension $d_b - d_c > 0$. Let $P_0$ be a projector onto this 
eigenspace and
\begin{equation}
    \tilde \om_b := P_0 \om_b P_0 = P_0 \sg_b P_0.
    \label{projEq}
\end{equation}
Since $\sg_b$ has rank $d_b$, the operator $\tilde \om_b$ is positive definite
on the support of $P_0$, thus $\tilde \om_b$ has $(d_b - d_c)$ strictly positive
eigenvalues $\{e_i\}$. Elementary results from perturbation theory~(for instance
see Sec.~5.2 in~\cite{SakuraiNapolitano17}) show that all $d_b - d_c$ zero
eigenvalues of $\rho_b(\ep)$ at $\ep = 0$ become non-zero for positive $\ep$,
and to leading order in $\ep$ these eigenvalues increase linearly such that the
$i^{th}$ such eigenvalue is simply $\ep e_i$. As a result $S_b(\ep)$ has
an $\ep \log$-singularity.

{\bf Non-additivity}: A $\log$-singularity based mechanism is responsible for
non-additivity~\eqref{nonAddEq} when $\BC_1$ and its complement $\CC_1$ are
channels defined by $J_1$ in~\eqref{qTritIso} and $\BC_2$, along with its
complement $\CC_2$, are defined an isometry $J_2:\HC_{a2} \mapsto \HC_{b2} \ot
\HC_{c2}$ of the form,
\begin{equation}
    J_2 \ket{0} = \ket{00}, \quad J_2 \ket{1} = \sqrt{1-p} \ket{10} + \sqrt{p} \ket{01}.
\end{equation}
Here $\BC_2$ represents a qubit amplitude damping channel with damping
probability $p$. We shall be interested in the parameter region $1/2 \leq p
\leq 1$ where $\BC_2$ is antidegradable and $\QC^{(1)}(\BC_2) = \QC(\BC_2) =
0$~\cite{WolfPerezGarcia07}. Consider the channel pair $\BC= \BC_1 \ot \BC_2,
\CC = \CC_1 \ot \CC_2$, with channel input
\begin{equation}
    \rho_a(\ep) = (1-w)[00]_a + w [\chi_{\ep}]_a, \quad
    \label{rhoChoice} 
\end{equation}
where 
\begin{equation}
    \ket{\chi_{\ep}}_a = \sqrt{1- \ep}\ket{10}_a + \sqrt{\ep} \ket{21}_a,
\end{equation}
$0 \leq \ep \leq 1$, and $w$ is chosen such that at $\ep = 0$,
\begin{equation}
    \rho_a(0) = \big((1-w)[0]_{a1} + w [1]_{a1} \big) \ot [0]_{a2} = \rho_{a1}^* \ot \rho_{a2}^*,
\end{equation}
i.e., $\Dl(0) = \QC^{(1)}(\BC_1) + \QC^{(1)}(\BC_2)$. For any $\ep>0$ and
$s>0$, an eigenvalue $\big((1-p)w\big) \ep$ of $\rho_b(\ep)$ and an eigenvalue
$(pkw)\ep$, 
\begin{equation}
    k = (1-s)(1-w)/\big(w + (1-s)(1-w) \big) < 1,
\end{equation}
of $\rho_c(\ep)$ increases linearly from zero to leading order in $\ep$. Thus
$S_b(\ep)$ has an $\ep \log$-singularity of rate $(1-p)w$ and $S_c(\ep)$ has a
$\ep \log$-singularity of rate $pkw$. The $\ep \log$-singularity in $S_b(\ep)$
is stronger when $p < \bar p(s) = 1/(1+k)$~(plotted in Fig.~\ref{FigD}).  This
stronger singularity implies non-additivity, i.e., a strict inequality in
eq.~\eqref{nonAddEq}.

\section*{Data availability}
No data sets were generated during this study.

\section*{Code availability}
Source code for the plots in this study are available on a Github repository,
\href{https://github.com/vsiddhu/logSing}{https://github.com/vsiddhu/logSing}.


\section*{Acknowledgments}
The author thanks Robert B. Griffiths, Graeme Smith, Costin B\u{a}descu, Yang Gao,
Michael Widom, and Mark M. Wilde for their useful comments. This work was
partially supported by NSF CAREER Award CCF 1652560 and NSF Grant PHY 1915407.

\section*{Author contributions}
The author performed research and wrote this paper.

\section*{Competing interests}
The author declares no competing interests.

\appendix

\section{Preliminaries}
\label{Aprlm}

Let $\HC$ denote a $d$-dimensional Hilbert space and $\hat \HC$ the space
of linear operators on $\HC$. The $l_1$ norm of any operator $A \in \hat \HC$,
$|A|_1:= \Tr(\sqrt{A A^{\dag}})$, defines an $l_1$ distance between two
such operators $A$ and $B$, $l_1(A,B)=|A-B|_1$.
Consider an isometry,
\begin{equation}
    J:\HC_a \mapsto \HC_b \ot \HC_c ; \quad J^{\dag}J = I_a
    \label{iso}
\end{equation}
that maps an input Hilbert space $\HC_a$~(of dimension $d_a$) to a subspace of
a pair of output spaces $\HC_b \ot \HC_c$. In the main text, we have used a
more concise notation $a,$ and $b \ot c$ to refer to $\HC_a$ and $\HC_b \ot
\HC_c$, respectively. For completeness, we repeat some material from the main
text. The isometry $J$ generates a noisy quantum channel pair $(\BC,\CC)$ with
superoperators,
\begin{equation}
    \BC(A) = \Tr_c(JAJ^\dag), \quad \text{and} \quad
    \CC(A) = \Tr_b(JAJ^{\dag}),
    \label{chanDef}
\end{equation}
that take any element $A$ of $\hat \HC_a$ to $\hat \HC_b$ and $\hat \HC_c$,
respectively. Both channels in this $(\BC,\CC)$ pair represent a completely
positive trace preserving map, and each channel may be referred to as the
complement of the other channel. The dimensions $d_b$ and $d_c$, of outputs
$\HC_b$ and $\HC_c$, respectively are the ranks of $\BC(I_a)$ and $\CC(I_a)$,
respectively. These are the smallest possible output dimensions required to
define $(\BC,\CC)$~(in the notation of Def.~4.4.4 in~\cite{Wilde17}, $d_b$ is
the Choi-rank of $\CC$ and $d_c$ is the Choi-rank of $\BC$). These definitions
make the channel pair setting symmetric with respect to replacement of one
channel in the pair with its complement.  In the main text, we sometimes write
$\BC: a \mapsto b$, to represent that the quantum channel $\BC$ takes operators
on $\HC_a$ to operators on $\HC_b$. 

Let $\rho$ be a density operator with eigenvalues $\{\lm_i\}$ then,
\begin{equation}
    S(\rho) = -\Tr(\rho \log \rho) = -\sum_i \lm_i \log \lm_i,
    \label{entVon}
\end{equation}
denotes the von-Neumann entropy of $\rho$. For
any input density operator $\rho_a$ in $\hat \HC_a$, $\rho_b$ and $\rho_c$
denote the outputs $\BC(\rho_a)$ and $\CC(\rho_a)$, respectively. The {\em
entropy bias} or the {\em coherent information} of $\BC$ at $\rho_a$ is 
\begin{equation}
    \Dl(\BC, \rho_a) = S(\rho_b) - S(\rho_c).
    \label{entBias}
\end{equation}
The {\em channel coherent information},
\begin{equation}
    \QC^{(1)}(\BC) = \max_{\rho_a} \Dl(\BC, \rho_a).
    \label{chanCI}
\end{equation}
If there is a channel $\DC$ such that $\CC = \DC \circ \BC$, then $\BC$ is said
to be {\em degradable} and $\QC(\BC) = \QC^{(1)}(\BC)$, $\CC$ is said to be
{\em antidegradable} and $\QC^{(1)}(\CC) = \QC(\CC) = 0$~\cite{DevetakShor05}. 

In what follow, we use the notation $[\psi]$ for the dyad $\dya{\psi}$.
First, in Appendix~\ref{ACnt}, we discuss the relationship between
log-singularities and continuity bounds. Next, in SN-3, we discuss properties of
channel outputs at rank 1 and full rank inputs. In SN-4, we discuss some
technical details about applications of Theorem~1 in the main text, where SN-4.3
uses some results from SN-3. Finally, SN-4 has mathematical details about
the isometry, $J_1$, and the channel coherent information, $\QC^{(1)}(\BC_1)$,
of the low-noise channel, $\BC_1$, all introduced in the main text.

\section{Continuity and $\log$-singularity}
\label{ACnt}

An $\ep \log$-singularity in the von-Neumann entropy can dominate continuity
bounds on this entropy. A simple continuity bound on the von-Neumann entropy
comes from the Fannes–Audenaert inequality~\cite{Fannes73, Audenaert07}. If the
$l_1$ distance between two $d$-dimensional density operators $\rho$ and $\sg$
is $2 \ep$, that is, $|\rho - \sg|_1 = 2 \ep$, then 
\begin{equation}
    |S(\rho) - S(\sg)| \leq \ep \log (d-1) + h(\ep),
    \label{eq:FanAu}
\end{equation}
where $h(\ep): = -[\ep \log_2 \ep + (1 - \ep) \log_2 (1-\ep)]$ is the binary
entropy function.
Suppose $\rho(\ep) = (1-\ep) [0] + \ep [1]$ is a qubit density operator where
$0 \leq \ep \leq 1$, and $\sg = \rho(0)$.  Then $\rho(\ep)$ and $\sg$ have
$l_1$ distance $2 \ep$ and satisfy~\eqref{eq:FanAu}. 
Notice, in the present case, the von-Neumann entropy of $\rho(\ep)$ has an $\ep
\log$-singularity and the right side of the inequality~\eqref{eq:FanAu} is
simply $h(\ep)$. The binary entropy function $h(\ep)$ has an $\ep
\log$-singularity in the sense that for small $\ep$, $d h(\ep)/d \ep \simeq
O\big( \log(\ep) \big)$; that is, the gradient of the upper
bound~\eqref{eq:FanAu} is logarithmic in $\ep$ and tends to infinity as $\ep$
tends to zero. In this sense, for small $\ep$, the continuity upper bound is
essentially dominated by a logarithmic singularity.

A continuity upper bound on the coherent information comes from the
Alick-Fannes-Winter~(AFW) inequality~\cite{AlickiFannes04, Winter16}. We find
that $\log$-singularities can also dominate the behavior of this bound. 
A bipartite density operator $\rho_{ab}$ on $\HC_{ab}:= \HC_a \ot \HC_b$ has
coherent information $I_{a>b}(\rho_{ab}) := S(\rho_{b}) - S(\rho_{ab})$.
The AFW inequality shows that if the $l_1$ distance between two density
operators $\rho_{ab}$ and $\sg_{ab}$ is at most $2 \ep$, $|\rho_{ab} -
\sg_{ab}|_1 \leq 2 \ep$, then coherent information of $\rho_{ab}$ and
$\sg_{ab}$ satisfy an inequality,
\begin{equation}
    |I_{a>b}(\rho_{ab}) - I_{a>b}(\sg_{ab})| \leq  2 \ep \log d_a + (1+ \ep) h\big( \frac{\ep}{1 + \ep}\big).
    \label{eq:AFW}
\end{equation}
Consider a density operator,
\begin{equation}
    \rho_{ab}(\ep) = (1-\ep) [00] + \ep [\phi],
    \label{eq:rhoEpCI}
\end{equation}
where $\ket{\phi}_{ab} = \frac{1}{\sqrt{2}} (\ket{01} + \ket{10})$, $\ket{ij}$
denotes $\ket{i}_a \ot \ket{j}_b \in \HC_a\ot \HC_b$, and let $\sg_{ab} =
\rho_{ab}(0)$. Then $\rho_{ab}(\ep)$ and $\sg_{ab}$ have $l_1$ distance $2 \ep$
and satisfy the AFW inequality~\eqref{eq:AFW}. The von-Neumann entropy of
$\rho_{ab}(\ep)$ has an $\ep \log$-singularity and for small $\ep$, changes in
$S\big( \rho_{ab} (\ep) \big)$ are logarithmic in $\ep$.  The upper bound
in~\eqref{eq:AFW} also experiences similar logarithmic changes, that is, for
small $\ep$ the gradient of the upper bound is $O(\log \ep)$ and this gradient
tends to infinity as $\ep$ tends to zero.

\section{Spectrum of channel outputs}
\label{AppAR1}
\subsection{Rank one inputs}
Suppose an input $\rho_a$ to a channel pair $(\BC, \CC)$~\eqref{chanDef} is a
normalized pure state $[\psi]_a$, then the channel outputs, $\rho_b$ and
$\rho_c$, have the same spectrum, rank, and entropy. This elementary result is
used at various places in the main text and this supplementary information
write-up.
To prove this result, we use the fact that two density operators with the same
spectrum have equal rank and entropy and show that $\rho_b$ and $\rho_c$ have
the same spectrum when $\rho_a = [\psi]_a$.
To obtain this spectrum consider the action of $J$ on the normalized
ket $\ket{\psi}_a$,
\begin{equation}
    J\ket{\psi}_a = \ket{\psi}_{bc} = \sum_i q_i \ket{\bt_i}_b \ot \ket{\gm_i}_c,
    \label{Schmidt}
\end{equation}
where $q_i>0$ are Schmidt coefficients with $\sum_i q^2_i =
1$, and $\{ \ket{\bt_i}_b \}$ and $\{ \ket{\gm_i}_c\}$ are orthonormal kets in
$\HC_b$ and $\HC_c$, respectively.
Using the Schmidt decomposition~\eqref{Schmidt} and eq.~\eqref{chanDef} we
obtain,
\begin{equation}
    \rho_b = \Tr_c([\psi]_{bc}) = \sum_i q^2_i [\bt_i]_b,
    \quad \text{and} \quad
    \rho_c = \Tr_b([\psi]_{bc})  = \sum_i q^2_i [\gm_i]_c,
\end{equation}
the spectral decompositions of $\rho_b$ and $\rho_c$, respectively. These
decompositions indicate that both channel outputs $\rho_b$ and $\rho_c$ have 
the same spectrum with eigenvalues $\{q^2_i\}$, thus proving our elementary 
result.

\subsection{Full rank inputs}
\label{AppARfull}

If a channel input $\rho_a$ has rank $d_a$, then the channel outputs, $\rho_b$
and $\rho_c$, have ranks $d_b$ and $d_c$, respectively. To prove this
statement, consider the spectral decomposition,
\begin{equation}
    \rho_a = \sum_{i=1}^{d_a} p_i [\al_i]_a,
    \label{lmA}
\end{equation}
of a rank $d_a$ density operator $\rho_a$, i.e., $\{p_i\}$ are $d_a$ strictly
positive eigenvalues that sum to unity and $\{\ket{\al_i}_a\}$ is an
orthonormal basis of $\HC_a$. Using this decomposition, $\rho_b$ can be
written as a convex combination of density operators $\BC([\al_i]_a)$,
\begin{equation}
    \rho_b = \sum_{i=1}^{d_a} p_i \BC([\al_i]_a).
    \label{lmB}
\end{equation}
For any ket $\ket{\phi}_b \in \HC_b$,
\begin{equation}
    \Tr\big( \rho_b [\phi]_b \big) = \sum_i p_i \Tr\big( \BC([\al_i]_a) [\phi]_b \big),
    \label{InpPhiB}
\end{equation}
is a convex sum of non-negative numbers $\Tr\big( \BC([\al_i]_a) [\phi]_b
\big)$. This convex sum is strictly positive if each $p_i$ is $1/d_a$ i.e.,
$\rho_a$ in \eqref{lmA} is $I_a/d_a$ and $\rho_b$ in \eqref{lmB} is
$\BC(I_a)/d_a$, a rank $d_b$ operator~(see discussion below
eq.~\eqref{chanDef}).  This strict positivity of 
$\Tr\big( \rho_b [\phi]_b \big)$ at $p_i = 1/d_a$ implies for arbitrary
$\ket{\phi}_b$, 
\begin{equation}
    \Tr \big( \BC([\al_i]_a) [\phi]_b \big) > 0,
    \label{nonNeg}
\end{equation}
for some $\ket{\al_i}_a$. Notice $\Tr\big( \rho_b [\phi]_b \big)$
in~\eqref{InpPhiB} is the sum of non-negative numbers. The equation above
implies that at least one of these numbers is strictly positive. Consequently,
\eqref{InpPhiB} is strictly positive for arbitrary $\ket{\phi}_b$. This strict
positivity implies $\rho_b$ is positive definite, i.e., $\rho_b$ has rank
$d_b$. 
A straightforward modification of the above reasoning shows that 
$\rho_c$ has rank $d_c$ when $\rho_a$ is rank $d_a$.

\section{Theorem applications}
\label{AThrmAps}

Proofs for various applications of Theorem~1 from the main text, restated below
for convenience, are given in this supplementary note.

\begin{theorem}
\label{Athrm}
    Assume $d_c<d_b$ and $\BC$ maps some pure state $[\psi]_a$ to an output
    $\BC([\psi]_a)$ of rank $d_c$, then $\QC^{(1)}(\BC) > 0$.
\end{theorem}

\subsection{Channel with equal output and environment dimension}
In the main text we argued how Theorem~1 applies to channels $\BC$ with $d_c
\geq d_b$. Here we give an explicit example where Theorem~1 applies when $d_c =
d_b = 3$.
Consider an isometry $K:\HC_a \mapsto \HC_b\ot \HC_c$ given by 
\begin{align}
    K \ket{0} &= \sqrt{1-p} \ket{00} + \sqrt{p} \ket{11}, 
    \label{Jex1} \\ \nonumber
    K \ket{1} &= \ket{21}, \\ \nonumber
    K \ket{2} &= \ket{12},
\end{align}
where $0 \leq p \leq 1$, $\{\ket{0}, \ket{1}, \ket{2}\}$ is the standard
orthonormal basis. 
At $p=0$ each channel in the $(\BC,\CC)$ pair defined by $J$ is antidegradable
and has zero quantum capacity. For all other values of $p$, Theorem~1 shows that
both channels in the pair $(\BC,\CC)$ have positive $\QC^{(1)}$. 
We apply this theorem to sub-channels of $\BC$ and $\CC$. 

First, consider $\BC'$, a sub-channel of $\BC$ obtained by restricting the
input of $\BC$ to a subspace spanned by $\{\ket{0}, \ket{1}\}$. This sub-channel
$\BC'$ satisfies the conditions of Theorem~1. At $p = 1$, $\BC'$ has a
two-dimensional output which is larger than its one dimensional environment,
and $\BC'$ maps a pure state input $[1]$ to a one-dimensional output. For
$0<p<1$, $\BC'$ has an output dimension $3$ which is larger than its
environment dimension $2$, and $\BC'$ maps $[0]$ to an output of rank $2$.

Next, consider $\tilde \CC$, a sub-channel of $\CC$ obtained by restricting the
input of $\CC$ to a subspace spanned by $\{\ket{0}, \ket{2}\}$. This
sub-channel $\CC'$ also satisfies the conditions of Theorem~1.
At $p = 1$, $\CC'$ has a two-dimensional output which is larger than its one
dimensional environment, and $\CC'$ maps a pure state input $[1]$ to a
one-dimensional output. For $0<p<1$, $\CC'$ has an output dimension $3$ which
is larger than its environment dimension $2$, and $\CC'$ maps $[0]$ to an
output of rank $2$.

\subsection{Incomplete Erasure Channels}
In the main text we introduced the incomplete erasure channel $\CC$. We claimed
that any zero quantum capacity qubit channel $\CC_1$ with a qubit environment
can be used to assist this incomplete erasure channel in sending quantum
information. We also asserted that our Theorem can be used to prove this claim.
In what follows we provide a systematic treatment to support our statements.
For completeness we introduce the generalized erasure channel
pair~\cite{SiddhuGriffiths20} with superoperators,
\begin{align}
    \BC(A) &= (1 - \lm) \BC_1(A) \bigoplus \lm \TC(A), \quad \text{and}
    \label{glnEra} \\ \nonumber
    \CC(A) &= (1 - \lm) \CC_1(A) \bigoplus \lm \IC(A),
\end{align}
where $(\BC_1,\CC_1)$ is an arbitrary channel pair, $\TC(A) = \Tr(A) [0]$ is
the trace channel, $\IC(A)=A$ is the identity channel, $\bigoplus$ is the
direct sum symbol, and $0 \leq \lm \leq 1$. Here $\CC$ is the incomplete
erasure channel. We are interested in the case where $(\BC_1,\CC_1)$ is some
qubit channel pair such that $\QC(\CC_1) = 0$. Up to local unitaries, any such
channel pair is generated by an isometry $K_1:\HC_a \mapsto \HC_{b1} \ot
\HC_{c1}$ of the form~\cite{WolfPerezGarcia07, RuskaiSzarekEA02}
\begin{align}
    K_1 \ket {0} &= \sqrt{1 -mp} \; \ket{00} + \sqrt{mp} \; \ket{11},
    \label{Jqubit} \\ \nonumber
    K_1 \ket {1} &= \sqrt{1 - p} \; \ket{10} + \sqrt{p} \; \ket{01},
\end{align}
where $0\leq m \leq 1$ and $0 \leq p \leq 1/2$ such that $\CC_1$ is
antidegradable and $\QC(\CC_1) = 0$. This zero capacity qubit channel $\CC_1$
has a qubit environment and a noise parameter $p$. The second channel parameter
$m$ can describe the type of noise, for instance at $m=0$, $\CC_1$ is an
amplitude damping channel, and at $m=1$ $\CC_1$ is a measure-and-prepare
channel~\cite{HorodeckiShorEA03}.

At $p=0$, $\CC_1$ is the trace channel $\TC$ and $\CC$ is an erasure channel
with erasure probability $1-\lm$. For this erasure channel both
$\QC^{(1)}(\CC)$ and $\QC(\CC)$ equal
$\max(0,2\lm-1)$~\cite{BennettDiVincenzoEA97} i.e., they are both zero for all
$0 \leq \lm \leq 1/2$. But as soon as $p$ is made positive by an arbitrarily
small amount, $\QC^{(1)}(\CC)$ becomes positive over the entire $\lm > 0$
range. This positivity comes from applying Theorem~1 to the
$\CC$ channel : $d_b = 3<d_c=4$ and a pure state $\ket{\psi}_a = (\ket{0}_a + i
\ket{1}_a)/\sqrt{2}$ is mapped to an output $\CC([\psi]_a)$ of rank $d_b$.

\subsection{Corollaries}
Next, we restate and prove Corollary~1 from the main text and show how
it can be applied to the complement qubit channels.

\begin{corollary}
\label{Acr1}
Suppose $d_a > 1$ and $d_b > d_a(d_c - 1)$ then $\QC^{(1)}(\BC)>0$.
\end{corollary}
\begin{proof}
    Follows from Theorem~1 by noting that if $d_a>1$ and $d_b >
    d_a(d_c - 1)$ then $d_c<d_b$ and there exists some pure state $[\psi]_a$ whose
    output, $\BC([\psi]_a)$, has rank $d_c$.
    The existence of such a pure state $[\psi]_a$ can be shown by contradiction
    as follows.
    Given $d_b>d_a(d_c-1)$ and $d_a>1$, assume no pure state $[\psi]_a$ has an
    output $\BC([\psi]_a)$ of rank $d_c$.  As discussed in the previous
    section, any pure state $[\psi]_a$ has outputs $\BC([\psi]_a)$ and
    $\CC([\psi]_a)$ of equal rank.  Consequently, this rank can never be
    greater than $\min (d_b,d_c) = d_c$, and by assumption this rank is not
    $d_c$, hence all pure states must be mapped by $\BC$ to outputs of rank at
    most $d_c -1$.  Given any orthonormal basis $\{\ket{\al_i}_a \}$ of
    $\HC_a$, any pure state $[\al_i]_a$ must get mapped by $\BC$ to an output
    $\BC([\al_i]_a)$ of rank at most $d_c - 1$.
    Consequently the sum of operators,
    \begin{equation}
        \sum_{i=1}^{d_a} \BC( [\al_i]_a ) = \BC \big( \sum_{i=1}^{d_a}  [\al_i]_a \big) = \BC(I_a),
        \label{sumEq}
    \end{equation}
    evaluated using linearity of $\BC$ and $I_a = \sum_i [\al_i]_a$, has rank
    at most $d_a(d_c-1)$. By definition~(see comments below \eqref{chanDef}),
    $\BC(I_a)$ has rank $d_b$, thus we arrive at an inequality $d_b \leq
    d_a(d_c - 1)$, which contradicts our starting condition $d_b > d_a(d_c -
    1)$. 
    Thus given that $d_b>d_a(d_c - 1)$ and $d_a>1$ one cannot assume that no
    pure state $[\psi]_a$ has an output $\BC([\psi]_a)$ of rank $d_c$ i.e.,
    there must be some pure state input $[\psi]_a$ with output $\BC([\psi]_a)$
    of rank $d_c$.

\end{proof}

In the main text, we claimed that Corollary~\ref{Acr1}, stated
above, shows that any qubit channel with a three of four dimensional
environment has a complement with non-zero quantum capacity. Now, we prove this
claim.
Replacing $\BC$ with its complement $\CC$ in Corollary~\ref{Acr1} results in the
following statement: if $d_a>1$ and $d_c > d_a(d_b - 1)$ then
$\QC^{(1)}(\CC)>0$. When $d_a = d_b = 2$, i.e., $\BC$ is a qubit channel, then
the above statement leads to
\begin{corollary}
    \label{cr2}
The complement $\CC$ of a qubit channel $\BC$ has strictly positive
$\QC^{(1)}(\CC)$ whenever $d_c>2$.
\end{corollary} 

\section{Qutrit Channel and Coherent information}
\label{AQCI}

We discuss properties of the isometry $J_1: \HC_{a1} \mapsto \HC_{b1} \ot
\HC_{c1}$ in eq.~(6) of the main text.  Below that equation, we stated that an
exchange of $s$ with $1-s$ can be achieved by local unitaries.  These unitaries
exchange $\ket{1}$ and $\ket{2}$ in $\HC_{a1}$, and exchange $\ket{0}$ and
$\ket{1}$ in both $\HC_{b1}$ and $\HC_{c1}$. The isometry $J_1$ gives rise to
the channel pair $(\BC_1,\CC_1)$. For the full range of $0 \leq s \leq 1/2$
values, $\QC^{(1)}(\BC_1)$ is strictly positive and is simply given by
$\Dl(\BC_1, \rho_{a1}^*)$ where,
\begin{equation}
    \rho_{a1}^* =(1-w) [0]_{a1} + w [1]_{a1},
    \label{rhoA2Star}
\end{equation}
and $0<w<1$. A proof of this previous statement is discussed below.
For the purposes of this discussion, we can get away with a more concise
notation than the one in the rest of this supplementary information write-up:
let $\rho \in \hat \HC_{a1}$ be any density operator with matrix elements
$\rho^{ij} = \bra{i}  \rho \ket{j}$ where $i,j \in \{0,1,2\}$, $S_b$ and $S_c$
be the von-Neumann entropies of $\BC_1(\rho)$ and $\CC_1(\rho)$, respectively,
and $\Dl := S_{b} - S_{c}$ be the entropy bias whose maximum over $\rho$ gives
the channel coherent information $\QC^{(1)}(\BC_1)$.

For any input density operator $\rho$ setting $\rho^{01}= \rho^{02} = 0$ 
increases $S_b$ without changing $S_{c}$, so we can always maximize $\Dl$ 
using a density operator of the form,
\begin{equation}
    \rho = \begin{pmatrix}
        1-w & 0 & 0 \\
        0 & w(1+z)/2 & w(x+iy)/2 \\
        0 & w(x+iy)/2 & w(1-z)/2 
    \end{pmatrix},
    \label{prm1}
\end{equation}
written in the standard basis using real parameters $0 \leq w \leq 1$ and
$x,y,$ and $z$ all between $-1$ and $1$ such that $x^2+y^2+z^2 \leq 1$.
In this parametrization, $S_{b}$ is independent of $x,y$, and $z$, on the other
hand $S_{c}$ depends on $x,y$ through $x^2+y^2$ so we set $y = 0$. Next, at
$y=0$, $S_{c}$ in concave in $x^2 + z^2$ thus the maximum value of $\Dl$ occurs
when $S_c$ is minimum at $x^2 + z^2 = 1$. These simplifications leave two free
parameters $0 \leq w \leq 1$ and $-1 \leq z \leq 1$. For a fixed $w$, at
$s=1/2$ the entropy $S_{c}$ is independent of $z$, and for other $0 \leq s <
1/2$, $S_c$ is monotone decreasing in $z$, as a consequence for all $0 \leq s
\leq 1/2$, we set $z = 1$ to maximize $\Dl$ at 
\begin{equation}
    \rho = (1 - w) [0] + w [1].
    \label{prm3}
\end{equation}
For any $0 \leq s \leq 1/2$, if we let $w$ in \eqref{prm3} be a small positive
number $\ep$, then the entropy $S_{b}$ has an $\ep \log$-singularity, while
$S_{c}$ doesn't, as a consequence $\QC^{(1)}(\BC_1) > 0$. Since
$\QC^{(1)}(\BC_1)$ is positive, its value can be obtained my maximizing $\Dl$
over $\rho$ in \eqref{prm3} by varying $w$ between zero and one while excluding
$w=0$ and $w=1$ where $\Dl = 0$.

\end{document}